# $B_c$ Meson Form factors and $B_c \to PV$ Decays involving Flavor Dependence of Transverse Quark Momentum

*Rohit Dhir and R.C. Verma*
Department of Physics,
Punjabi University,
Patiala-147002, INDIA.

## Abstract

We present a detailed analysis of the $B_c$ form factors in the BSW framework, by investigating the effects of the flavor dependence on the average transverse quark momentum inside a meson. Branching ratios of two body decays of $B_c$ meson to pseudoscalar and vector mesons are predicted.

## 1. Introduction

The discovery of the $B_c$ meson by the collider detector at Fermilab (CDF) [1] opens up some interesting investigations concerning the structure of strong and weak interactions. The properties of the $B_c$ meson are of special interest [2], since it is the only heavy meson consisting of two heavy quarks with different flavors. This difference of quarks flavor forbids annihilation in to gluons. A peculiarity of the $B_c$ decays, with respect to the decays of $B$ and $B_s$ mesons, is that both the quarks may involve in its weak decays. There are quite a few numbers of theoretical works studying various leptonic, semileptonic and hadronic decay channels of $B_c$ mesons in different models [3-14]. Their estimates of $B_c$ decay rates indicate that the $c$-quark give dominant contribution as compared to $b$-quark decays. From experimental point of view, study of weak decays of $B_c$ meson is quite important for the determination of CKM elements. More detailed information about its decay properties are expected in the near future at LHC and other experiments.

In our recent work [14], we have investigated the effects of flavor dependence on $B_c \to P$ form factors, caused by possible variation of average transverse quark momentum ($\omega$) in a meson. Employing the BSW framework [15], we then predicted the branching ratios of $B_c$ meson decaying to two pseudoscalar mesons. In the present paper, we extend our analysis to investigate such effects on the form factors involving $B_c \to V$ transitions. We also calculate the branching ratios of $B_c$ meson decaying to a pseudoscalar (P) meson and a vector (V) meson. We observe that the branching ratios of $B_c$ decays get enhanced for both the bottom changing and bottom conserving decay modes of $B_c$ meson, when such flavor dependent effects are included.

The present paper is organized as follows: In Section 2, we give the methodology. Section 3 deals with $B_c$ form factors in the BSW model. We study the effects of flavor dependence of $\omega$ on $B_c \to V$ form factors in Section 4. Finally the branching ratios of $B_c \to PV$ decays are predicted. Section 5 contains summary and conclusions.

## 2. Methodology

The decay rate is given by

$$\Gamma(B_c \to PV) = \frac{k^3}{8\pi m_V^2} |A(B_c \to PV)|^2, \qquad (1)$$

where in the three-momentum $k$ of the final state particle in the rest frame of $B_c$ is given by

$$k = \frac{1}{2m_{B_c}} \{[m_{B_c}^2 - (m_P + m_V)^2][m_{B_c}^2 - (m_P - m_V)^2]\}^{1/2}. \qquad (2)$$

## 2.1 Weak Hamiltonian

The QCD modified weak Hamiltonian generating [16] the *b*-quark decays in CKM enhanced modes ($\Delta b = 1, \Delta C = 1, \Delta S = 0; \Delta b = 1, \Delta C = 0, \Delta S = -1$) is given by

$$H_w^{\Delta b=1} = \frac{G_F}{\sqrt{2}} \{V_{cb}V_{ud}^*[c_1(\mu)(\bar{c}b)(\bar{d}u) + c_2(\mu)(\bar{c}u)(\bar{d}b)] + V_{cb}V_{cs}^*[c_1(\mu)(\bar{c}b)(\bar{s}c) + c_2(\mu)(\bar{c}c)(\bar{s}b)]\}, \qquad (3)$$

where $\bar{q}q \equiv \bar{q}\gamma_\mu(1-\gamma_5)q$, $G_F$ is the Fermi constant and $V_{ij}$ are the CKM matrix elements, $c_1$ and $c_2$ are the standard perturbative QCD coefficients.

In addition to the bottom changing decays, bottom conserving decay channel is also available for the $B_c$ meson, where the charm quark decays to an *s* or *d* quark. The weak Hamiltonian generating the *c*-quark decays in CKM enhanced mode ($\Delta b = 0, \Delta C = -1, \Delta S = -1$) is given by

$$H_w^{\Delta c=-1} = \frac{G_F}{\sqrt{2}} V_{ud}V_{cs}^*[c_1(\mu)(\bar{u}d)(\bar{s}c) + c_2(\mu)(\bar{u}c)(\bar{s}d)]. \qquad (4)$$

One naively expects this channel to be suppressed kinematically due to the small phase space available. However, the kinematic suppression is well compensated by the CKM element $V_{cs}$, which is larger than $V_{cb}$ appearing in the bottom changing decays [13]. In fact, we shall show later that bottom conserving decay modes are more prominent than the bottom changing ones.

## 2.2 Factorization scheme

In the standard factorization scheme, the decay amplitude is obtained by sandwiching the QCD modified weak Hamiltonian which is given below:

$$A(B_c \to PV) \propto \langle P|J^\mu|0\rangle\langle V|J_\mu^\dagger|B_c\rangle + \langle V|J^\mu|0\rangle\langle P|J_\mu^\dagger|B_c\rangle, \qquad (5)$$

where the weak current $J_\mu$ is given by

$$J_\mu = (\bar{u}\ \bar{c}\ \bar{t})\gamma_\mu(1-\gamma_5)\begin{pmatrix} d' \\ s' \\ b' \end{pmatrix}, \qquad (6)$$

and $d'$, $s'$, $b'$ are mixture of the $d$, $s$ and $b$ quarks, as given by the Cabibbo-Kobayashi-Maskawa (CKM) matrix [17].

Matrix elements of the currents are defined [15] as,

$$\langle V|J_\mu|B_c\rangle = \frac{2}{m_{B_c}+m_V}\varepsilon_{\mu\nu\rho\sigma}\varepsilon^{*\nu}P_{B_c}^\rho P_V^\sigma V(q^2)$$
$$+i\{\varepsilon_\mu^*(m_{B_c}+m_V)A_1(q^2) - \frac{\varepsilon^*\cdot q}{m_{B_c}+m_V}(P_{B_c}+P_V)_\mu A_2(q^2) \quad (7)$$
$$-\frac{\varepsilon^*\cdot q}{q^2}2m_V q_\mu A_3(q^2)\} + i\frac{\varepsilon^*\cdot q}{q^2}2m_V q_\mu A_0(q^2),$$

$$\langle P|J_\mu|B_c\rangle = (P_{B_c}+P_P - \frac{m_{B_c}^2-m_P^2}{q^2}q)_\mu F_1(q^2) + \frac{m_{B_c}^2-m_P^2}{q^2}q_\mu F_0(q^2), \quad (8)$$

$$\langle P|J_\mu|0\rangle = -i\,f_P P_\mu, \quad (9)$$
$$\langle V|J_\mu|0\rangle = \epsilon_\mu^* f_V m_V, \quad (10)$$

where $\varepsilon_\mu$ denotes the polarization vector of the outgoing vector meson $q_\mu = (P_{B_c}-P_P)_\mu$, $F_1(0)=F_0(0)$, $A_3(0)=A_0(0)$ and

$$A_3(q^2) = \frac{m_{B_c}+m_V}{2m_V}A_1(q^2) - \frac{m_{B_c}-m_V}{2m_V}A_2(q^2). \quad (11)$$

There are three types of $B_c$ decays:

(i) caused by color favored diagram,
(ii) caused by color suppressed diagram, and
(iii) caused by both color favored and color suppressed diagrams.

In general, the color favored decay amplitude $A(B_c \to PV)$ can be expressed as

$$A(B_c \to PV) = \frac{G_F}{\sqrt{2}}\times(CKM\ factors)\times 2m_V\,a_1 \quad (12)$$
$$\times\{(C.G.Coeff.)\,f_V\,F_1^{B_cP}(m_V^2) + (C.G.Coeff.)\,f_P\,A_0^{B_cV}(m_P^2)\}$$

where [15] $a_1(\mu) = c_1(\mu) + \frac{1}{N_c}c_2(\mu)$, and $N_c$ is the number of colors. For the color suppressed modes, QCD factor $a_1$ is replaced by $a_2$ which is given as $a_2(\mu) = c_2(\mu) + \frac{1}{N_c} c_1(\mu)$. However, we follow the convention of large $N_c$ limit to fix QCD coefficients $a_1 \approx c_1$ and $a_2 \approx c_2$, where [16]:

$$c_1(\mu) = 1.26 \ , \ c_2(\mu) = -0.51 \text{ at } \mu \approx m_c^2,$$
$$c_1(\mu) = 1.12 \ , \ c_2(\mu) = -0.26 \text{ at } \mu \approx m_b^2. \quad (13)$$

To evaluate the factorization amplitudes (9) and (10), we use the following decay constants [5, 8, 17]:

$$f_\pi = 0.131 \text{ GeV}, \ f_K = 0.160 \text{ GeV}, \ f_D = 0.208 \text{ GeV},$$
$$f_{D_s} = 0.273 \text{ GeV}, \ f_{\eta_c} = 0.400 \text{ GeV},$$

and

$$f_\rho = 0.221 \text{ GeV}, \ f_{K^*} = 0.220 \text{ GeV}, \ f_{D^*} = 0.245 \text{ GeV},$$
$$f_{D^*_s} = 0.273 \text{ GeV}, \ f_{J/\psi} = 0.411 \text{GeV}. \quad (14)$$

It has been pointed out in the BSW2 model [18] that consistency with the heavy quark symmetry requires certain form factors such as $F_1$ and $A_0$ to have dipole $q^2$ dependence i.e.
$$F_1(q^2) = F_1(0)/(1-q^2/m_V^2)^2 \text{ and } A_0(q^2) = A_0(0)/(1-q^2/m_P^2)^2.$$
Therefore, in this work, we have determined the amplitudes using the dipole $q^2$ dependence for these form factors.

**3. Form factors in BSW model**

We employ the BSW model for evaluating the meson form factors. In this model, the initial and final state mesons are given by the relativistic bound states of a quark $q_1$ and an antiquark $\bar{q}_2$ in the infinite momentum frame [15],

$$|\mathbf{P}, m, j, j_z\rangle = \sqrt{2}(2\pi)^{3/2} \sum_{s_1 s_2} \int d^3 p_1 d^3 p_2 \delta^3(\mathbf{P} - \mathbf{p_1} - \mathbf{p_2})$$
$$\times \psi_m^{j,j_z}(\mathbf{p_{1T}}, x, s_1, s_2) a_1^{S_1^+}(\mathbf{p_1}) b_2^{S_2^+}(\mathbf{p_2})|0\rangle , \quad (15)$$

where $P_\mu = (P_0, 0, 0, P)$ with $P \to \infty$, $x$ denotes the fraction of the longitudinal momentum carried by the non-spectator quark $q_1$, and $\mathbf{p_{1T}}$ denotes its transverse momentum:

$$x = p_{1Z}/P, \ \mathbf{p_{1T}} = (p_{1x}, p_{1y}).$$

Though $B_c \to PV$ decays involve $F_1(q^2)$ and $A_0(q^2)$ only, we calculate all the form factors appearing in the expression (7) and (8) to later investigate their flavor dependence.

By expressing the current $J_\mu$ in terms of the annihilation and creation operators, the form factors are given by the following integrals:

$$F_0^{B_c P}(0) = F_1^{B_c P}(0) = \int d^2 p_T \int_0^1 (\psi_P^*(\mathbf{p_T}, x)\psi_{B_c}(\mathbf{p_T}, x))dx,$$

$$A_0^{B_c V}(0) = A_3^{B_c V}(0) = \int d^2 \mathbf{p_T} \int_0^1 dx(\psi_V^{*1,0}(\mathbf{p_T}, x)\sigma_Z^{(1)}\psi_{B_c}(\mathbf{p_T}, x)), \quad (16)$$

$$V(0) = \frac{m_{q_1(B_c)} - m_{q_1(V)}}{m_{B_c} - m_V} I, \quad (17)$$

and

$$A_1(0) = \frac{m_{q_1(B_c)} + m_{q_1(V)}}{m_{B_c} + m_V} I, \quad (18)$$

where

$$I = \sqrt{2}\int d^2 \mathbf{p_T} \int_0^1 \frac{dx}{x}(\psi_V^{*1,-1}(\mathbf{p_T}, x)i\sigma_y^{(1)}\psi_{B_c}(\mathbf{p_T}, x)), \quad (19)$$

$m_{q_1(B_c)}$ and $m_{q_1(V)}$ denote masses of the non-spectator quarks participating in the quark decay process. Meson wave function is given by

$$\psi_m(\mathbf{p_T}, x) = N_m\sqrt{x(1-x)}\exp(-\mathbf{p_T}^2/2\omega^2)\exp(-\frac{m^2}{2\omega^2}(x - \frac{1}{2} - \frac{m_{q_1}^2 - m_{q_2}^2}{2m^2})^2), \quad (20)$$

where $m$ denotes the meson mass and $m_i$ denotes the $i^{th}$ quark mass, $N_m$ is the normalization factor and $\omega$ is the average transverse quark momentum, $\langle \mathbf{p_T}^2 \rangle = \omega^2$.

The form factors are sensitive to the choice of $\omega$, which is treated as a free parameter. In the BSW model [15], the form factors are calculated by taking $\omega = 0.40$ GeV for all the mesons and $m_u = m_d = 0.35$ GeV, $m_s = 0.55$ GeV, $m_c = 1.7$ GeV and $m_b = 4.9$ GeV. The $B_c \to P$ form factors thus obtained are given in column 3 of Table I and $B_c \to V$ form factors are given in Table II. Using them, we obtain the branching ratios for various $B_c$ decays, with $\tau_{B_c} = 0.46\,\text{ps}$ as given in column 2 of Table IV. We make the following observations:

1. Naively, one may expect the bottom conserving modes to be kinematically suppressed. However, the large CKM mixing angle for bottom conserving modes overcomes this suppression. We find that the bottom conserving and charm changing modes are dominant: $B(B_c^+ \to \pi^+ B_s^{*0}) = 1.91\%$, $B(B_c^+ \to B_s^0 \rho^+) = 2.75\%$, $B(B_c^+ \to B^+ \overline{K}^{*0}) = 0.38\%$, and $B(B_c^+ \to \overline{K}^0 B^{*+}) = 0.38\%$. Among the bottom changing decays $B(B_c^- \to \eta_c D_s^{*-}) = 0.04\%$, $B(B_c^- \to D_s^- J/\psi) = 0.03\%$, and $B(B_c^- \to \eta_c \rho^-) = 0.04\%$ modes dominate, but are small as compared to the bottom conserving modes.

2. Because of the less overlap of the initial and final state wave functions for $\omega = 0.40$ GeV, as shown in Fig. I, bottom changing modes are further suppressed due to the small values of the corresponding form factors.

3. Besides $B_c^- \to \eta_c D_s^{*-}$ and $B_c^- \to D_s^- J/\psi$ decays, various decays such as $B_c^- \to K^- \overline{D}^{*0}$, $B_c^- \to \eta' D_s^{*-}$, $B_c^- \to \overline{D}^0 K^{*-}$, $B_c^- \to D_s^- \rho^0$, $B_c^- \to D_s^- \omega$, $B_c^- \to D_s^- \phi$, $B_c^- \to \pi^0 D_s^{*-}$, and $B_c^- \to \eta D_s^{*-}$ are also permitted by the selection rule $\Delta b = 1$, $\Delta C = 0$, $\Delta S = -1$. However, their branching ratios are heavily suppressed as they occur through the CKM suppressed weak process involving $b \to u$ transitions.

## 4. Effects of flavor dependence on $B_c \to V/P$ form factors

In the previous work [14], we have investigated the possible flavor dependence in $B_c \to P$ form factors and consequently in $B_c \to PP$ decay widths. We wish to point out that the parameter $\omega$, being dimensional quantity, may show flavor dependence. Therefore, it may not be justified to take the same $\omega$ for all the mesons. Following the analysis described in [14], we estimate $\omega$ for different mesons from $|\psi(0)|^2$ i.e. square of the wave function at origin, using the following relation based on the dimensionality arguments

$$|\psi(0)|^2 \propto \omega^3. \tag{21}$$

$|\psi(0)|^2$ is obtained from the hyperfine splitting term for the meson masses [19],

$$|\psi(0)|^2 = \frac{9 m_i m_j}{32 \alpha_s \pi}(m_V - m_P), \tag{22}$$

where $m_V$ and $m_P$ respectively denotes masses of vector and pseudoscalar mesons composed of $i$ and $j$ quarks. The meson masses fix quark masses (in GeV) to be $m_u = m_d = 0.31$, $m_s = 0.49$, $m_c = 1.7$, and $m_b = 5.0$ for $\alpha_s(m_b) = 0.19$, $\alpha_s(m_c) = 0.25$, and $\alpha_s = 0.48$ (for light flavors $u$, $d$ and $s$).

Except for $B_c^*$, all the meson masses required are available experimentally. Theoretical estimates for hyperfine splitting $m_{B_c^*} - m_{B_c}$ obtained in different quark models [20, 21] range from 65 to 90 MeV. For the present work, we take $m_{B_c^*} - m_{B_c} = 73 \pm 15$ MeV obtained in [20], which has been quite successful in giving charmonium and bottomium mass spectra. Calculated numerical values of $|\psi(0)|^2$ are listed in column 2 of Table V. Variation in $\omega_{B_c}$ with hyperfine splitting $m_{B_c^*} - m_{B_c}$ is shown in the Fig. II. We use the well measured form factor $F_0^{DK}(0) = 0.78 \pm 0.04$ to determine $\omega_D = 0.43$ GeV which in turn yields $\omega$ for other mesons given in column 3 of Table II. Obtained form factors for $B_c \to P$ transitions are given in column 4 of Table I and for $B_c \to V$ transitions are given in Table III. We find that all the form factors get significantly enhanced due to the large overlap of $B_c$ and the final state meson as shown in Fig. III. In Fig. IV and Fig. V, we show the dependence of various $B_c \to B^*$ and $B_c \to D^*$ form factors on $\omega_{B_c}$ and $\omega_F$ in the range 0 to 1, where $\omega_F$ is for the final state meson.

**4.1 Numerical branching ratios**

Using the flavor dependent form factors, we finally predict branching ratios which are given in column 3 of Table IV. We observe the following:

1. The branching ratios get enhanced significantly for both bottom changing as well as for bottom conserving modes. However, the bottom conserving and charm changing mode still remain dominant with $B(B_c^+ \to \pi^+ B_s^{*0}) = 4.37^{+0.37}_{-0.33}\%$, $B(B_c^+ \to B_s^0 \rho^+) = 7.00^{+0.60}_{-0.50}\%$, $B(B_c^+ \to \bar{K}^0 B^{*+}) = 0.80^{+0.00}_{-0.14}\%$ and $B(B_c^+ \to B^+ \bar{K}^{*0}) = 0.72^{+0.15}_{-0.00}\%$.

2. Among the bottom changing modes, higher branching ratios are $B(B_c^- \to D_s^- J/\psi) = 0.28^{+0.01}_{-0.01}\%$, $B(B_c^- \to \eta_c D_s^{*-}) = 0.31^{+0.01}_{-0.01}\%$, $B(B_c^- \to \pi^- J/\psi) = 0.13^{+0.01}_{-0.01}\%$ and $B(B_c^- \to \eta_c \rho^-) = 0.39^{+0.01}_{-0.03}\%$. For the sake of the comparison, we list results of other models in Table IV for branching ratios of $B_c$ meson.

3. It may be noted that the decay widths of $B_c^- \to \eta_c D_s^{*-}$ and $B_c^- \to D_s^- J/\psi$ involve contributions from both the color favored and the color suppressed diagrams. In $B$ meson decays, the experimental data favors constructive interference [16], in contrast to the charm meson sector, between the color-favored and color suppressed diagrams, thereby yielding $a_1 = 1.10 \pm 0.08$ and $a_2 = 0.20 \pm 0.02$. Taking $a_1 = 1.10$ and $a_2 = 0.20$ for the constructive interference case, we obtain larger values, $B(B_c^- \to D_s^- J/\psi)$ = $0.49^{+0.02}_{-0.03}\%$ and $B(B_c^- \to \eta_c D_s^{*-})$ = $0.51^{+0.03}_{-0.03}\%$ in comparison to $0.28^{+0.01}_{-0.01}\%$ and $0.31^{+0.01}_{-0.01}\%$ respectively obtained for the destructive interference.

## 5. Summary and conclusions

In this paper, we have employed the BSW relativistic quark model to study the hadronic weak decays of $B_c$ meson decaying to a pseudoscalar and a vector meson in CKM enhanced mode. We have also investigated the flavor dependence of $\omega$, hence consequently of the form factors and the branching ratios for bottom changing as well as bottom conserving decay modes. We draw the following conclusions.

1. One naively expects the bottom conserving modes to be kinematically suppressed, however, the large CKM angle involved overly compensates the suppression. Due to the less overlap of the initial and the final state wave functions, the form factors involving the bottom changing transitions are small as compared to those of the bottom conserving transitions. As a result of which bottom changing decays get suppressed in comparison to bottom conserving decays.

2. Initially, form factors for both the modes are obtained by taking the usual value of $\omega = 0.40$ GeV for all the mesons. For bottom conserving and charm changing modes, their form factors yield $B(B_c^+ \to \pi^+ B_s^{*0}) = 1.91\%$, $B(B_c^+ \to B_s^0 \rho^+) = 2.75\%$, $B(B_c^+ \to B^+ \overline{K}^{*0}) = 0.38\%$, and $B(B_c^+ \to \overline{K}^0 B^{*+}) = 0.38\%$. Among the bottom changing decays, the dominant branching ratios are: $B(B_c^- \to D_s^- J/\psi) = 0.03\%$, $B(B_c^- \to \eta_c D_s^{*-}) = 0.04\%$ and $B(B_c^- \to \eta_c \rho^-) = 0.04\%$.

3. We then investigate the effects of possible flavor dependence of $\omega$. Determining $|\psi(0)|^2$ from the meson masses to fix $\omega$ for each meson, we calculate various form factors for $B_c$ transitions which get significantly enhanced for bottom changing as well as for bottom conserving transitions. However, bottom conserving decays remain dominant with higher branching ratios: $B(B_c^+ \to \pi^+ B_s^{*0}) = 4.37^{+0.37}_{-0.33}\%$, $B(B_c^+ \to B_s^0 \rho^+) = 7.00^{+0.60}_{-0.50}\%$, $B(B_c^+ \to \overline{K}^0 B^{*+}) = 0.80^{+0.00}_{-0.14}\%$ and $B(B_c^+ \to B^+ \overline{K}^{*0}) = 0.72^{+0.15}_{-0.00}\%$, while branching ratios of bottom changing modes are also increased to $B(B_c^- \to D_s^- J/\psi) = 0.28^{+0.01}_{-0.01}\%$, $B(B_c^- \to \eta_c D_s^{*-}) = 0.31^{+0.01}_{-0.01}\%$, $B(B_c^- \to \pi^- J/\psi) = 0.13^{+0.01}_{-0.01}\%$ and $B(B_c^- \to \eta_c \rho^-) = 0.39^{+0.01}_{-0.03}\%$. Errors in the predictions are due to uncertainty in theoretical estimate of hyperfine splitting $m_{B_c^*} - m_{B_c} = 73 \pm 15$ MeV.

4. Taking into account the constructive interference observed for $B$ meson decays involving both the color favored and color suppressed diagrams [15], we find that $B(B_c^- \to D_s^- J/\psi)$ and $B(B_c^- \to \eta_c D_s^{*-})$ gets further enhanced to $0.49^{+0.02}_{-0.03}\%$ and $0.51^{+0.03}_{-0.03}\%$ respectively. From the experimental point of view, measurements of these branching ratios present an interesting test for the interference between color favored and color suppressed processes in $B_c$ meson decays.

**Figure captions**

**Fig. I.**  Overlap of wave functions for $B_c \to \overline{D}^*$ decays at $\omega_{B_c} = \omega_{D^*} = 0.40$ GeV.

**Fig. II.**  Variation in $\omega_{B_c}$ with hyperfine splitting $m_{B_c^*} - m_{B_c}$.

**Fig. III.**  Overlap of wave functions for $B_c \to \overline{D}^*$ decays at $\omega_{B_c} = 0.96$ GeV and $\omega_{D^*} = 0.43$ GeV.

**Fig. IV.**  Variation of $B_c \to \overline{D}^*$ form factors $V(0)$, $A_0(0)$, $A_1(0)$ and $A_2(0)$ with $\omega$'s for initial and final state mesons.

**Fig. V.**  Variation of $B_c \to B^*$ form factors $V(0)$, $A_0(0)$, $A_1(0)$ and $A_2(0)$ with $\omega$'s for initial and final state mesons.

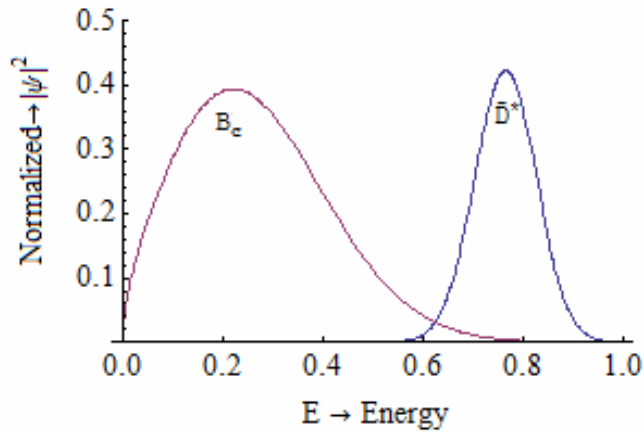

**Fig. I.**  Overlap of wave functions for $B_c \to \overline{D}^*$ decays at $\omega_{B_c} = \omega_{D^*} = 0.40$ GeV.

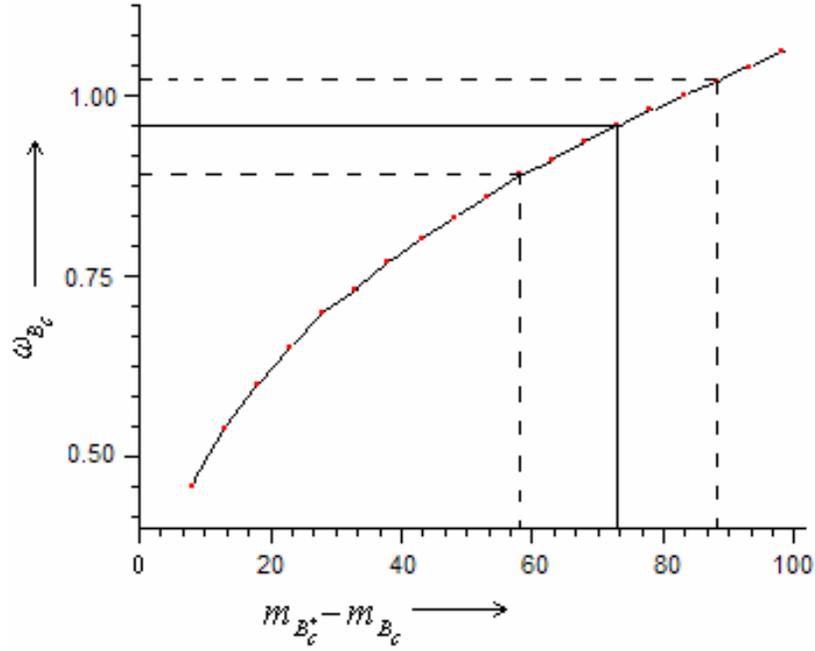

**Fig. II.** Variation in $\omega_{B_c}$ with hyperfine splitting $m_{B_c^*} - m_{B_c}$.

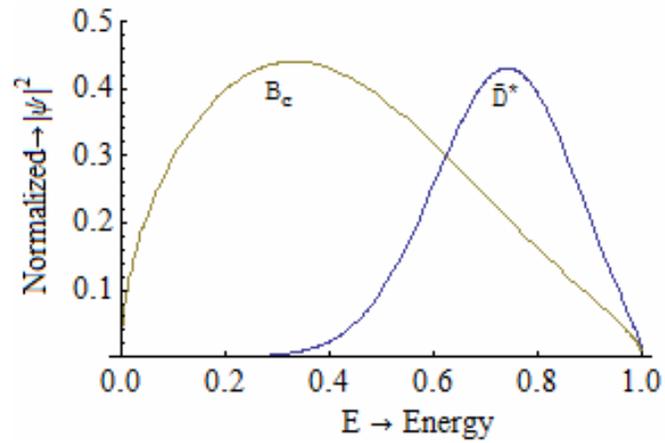

**Fig. III.** Overlap of wave functions for $B_c \to \overline{D}^*$ decays at $\omega_{B_c}$ = 0.96 GeV and $\omega_{D^*}$ = 0.43 GeV.

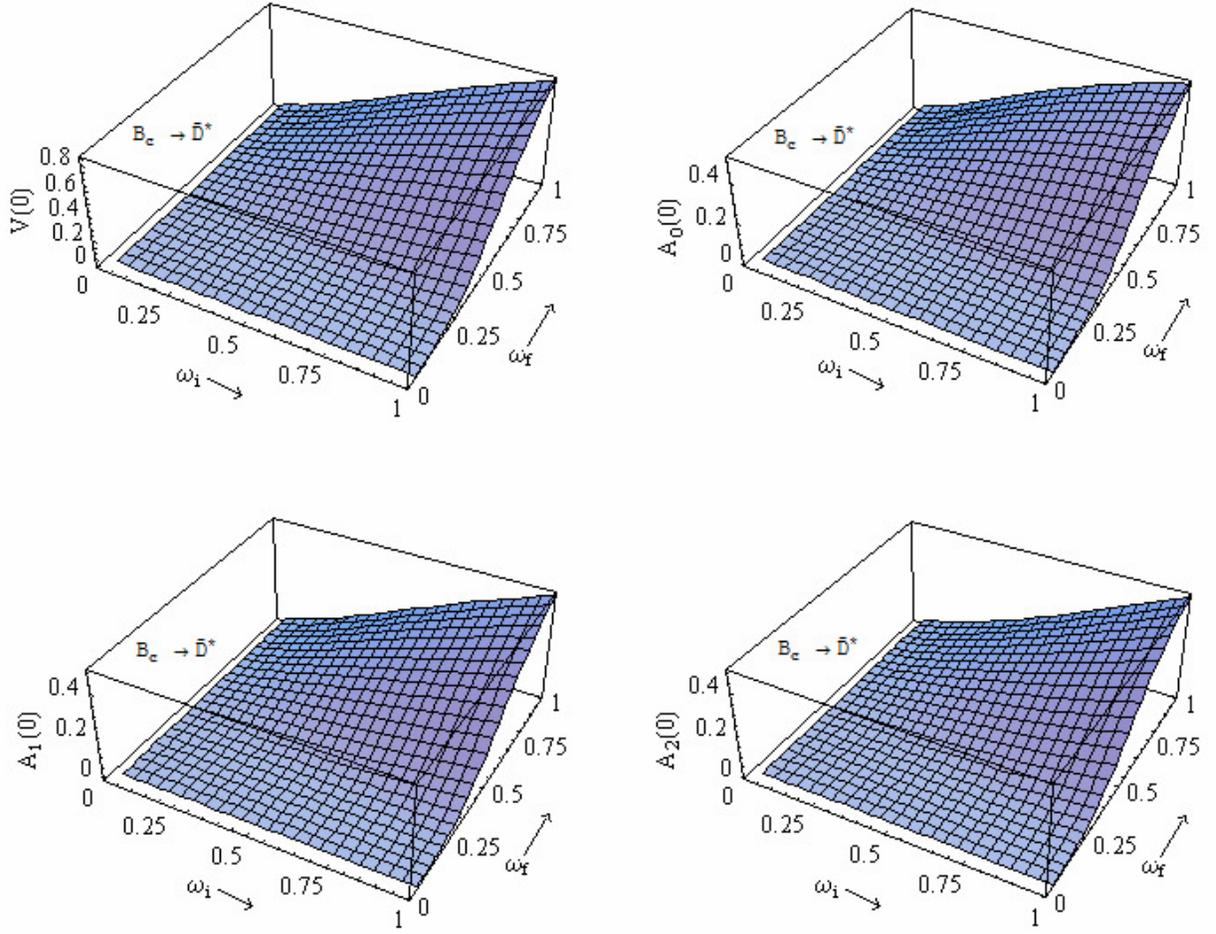

**Fig. IV.** Variation of $B_c \to \bar{D}^*$ form factors $V(0)$, $A_0(0)$, $A_1(0)$ and $A_2(0)$ with $\omega$'s for initial and final state mesons.

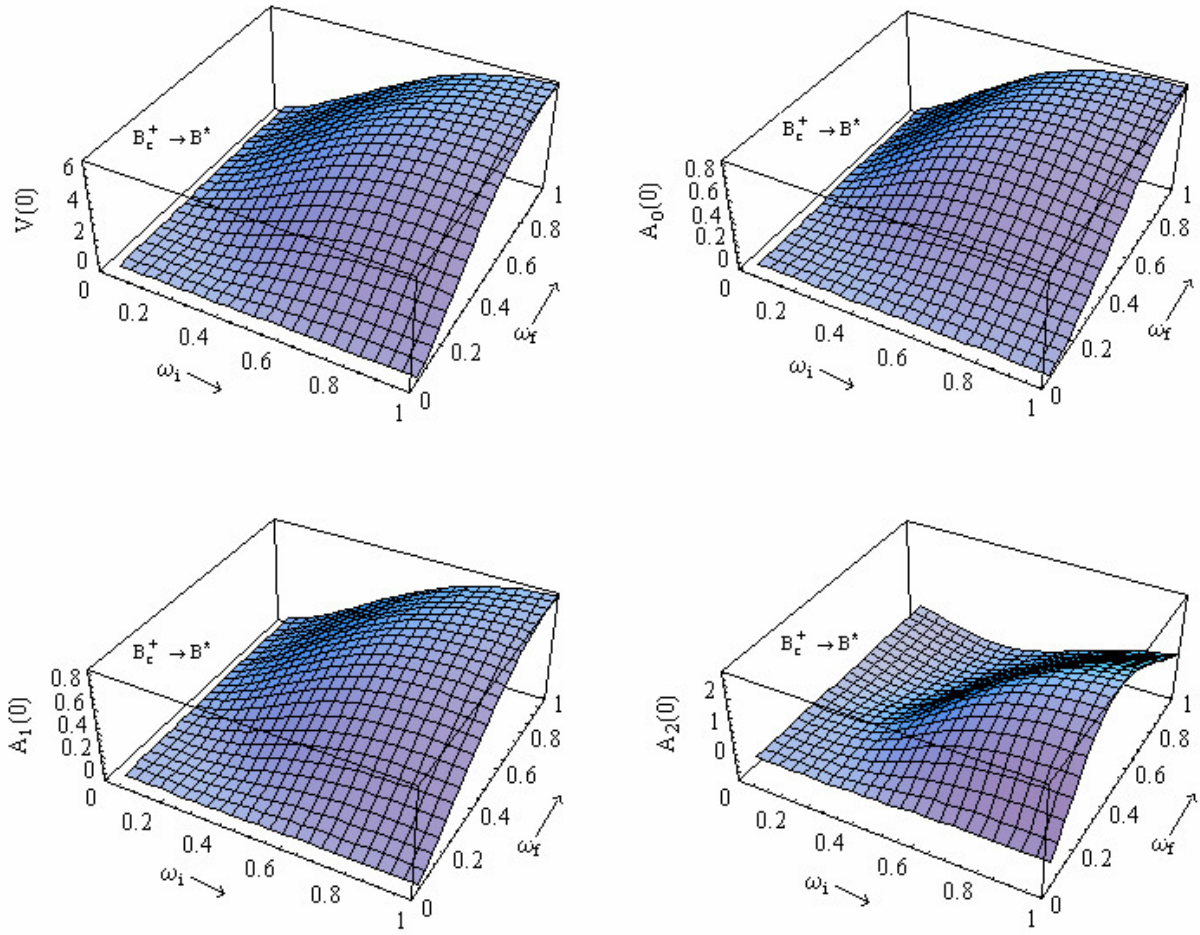

**Fig. V.** Variation of $B_c \to B^*$ form factors $V(0)$, $A_0(0)$, $A_1(0)$ and $A_2(0)$ with $\omega$'s for initial and final state mesons.

**Tables I.** Form factors of $B_c \to P$ transition (Errors shown here are due to uncertainty in $m_{B_c^*} - m_{B_c}$).

| Modes | Transition | This work $F_1^{B_c P}(0)$ ($\omega = 0.40$ GeV) | $F_1^{B_c P}(0)$ (using flavor dependent $\omega$) |
|---|---|---|---|
| $\Delta b = 0, \Delta C = -1, \Delta S = -1$ | $B_c \to B_s$ | 0.35 | $0.55^{+0.02}_{-0.02}$ |
| | $B_c \to B$ | 0.28 | $0.41^{+0.01}_{-0.02}$ |
| $\Delta b = 1, \Delta C = 0, \Delta S = -1$ | $B_c \to D$ | 0.015 | $0.075^{+0.06}_{-0.08}$ |
| | $B_c \to D_s$ | 0.021 | $0.15^{+0.01}_{-0.01}$ |
| $\Delta b = 1, \Delta C = 1, \Delta S = 0$ | $B_c \to \eta_c(c\bar{c})$ | 0.19 | $0.58^{+0.02}_{-0.01}$ |

**Tables II.** Form factors of $B_c \to V$ transition ($\omega = 0.40$ GeV)

| Modes | Transition | $V(0)$ | $A_0(0)$ | $A_1(0)$ | $A_2(0)$ |
|---|---|---|---|---|---|
| $\Delta b = 0, \Delta C = -1, \Delta S = -1$ | $B_c \to B_s^*$ | 2.45 | 0.37 | 0.40 | 0.68 |
| | $B_c \to B^*$ | 2.23 | 0.31 | 0.31 | 0.35 |
| $\Delta b = 1, \Delta C = 0, \Delta S = -1$ | $B_c \to D^*$ | 0.025 | 0.016 | 0.015 | 0.013 |
| | $B_c \to D_s^*$ | 0.032 | 0.022 | 0.020 | 0.019 |
| $\Delta b = 1, \Delta C = 1, \Delta S = 0$ | $B_c \to J/\psi(c\bar{c})$ | 0.24 | 0.17 | 0.17 | 0.17 |

**Tables III.** Form factors of $B_c \to V$ transition using flavor dependent $\omega$ (Errors shown here are due to uncertainty in $m_{B_c^*} - m_{B_c}$).

| Modes | Transition | $V(0)$ | $A_0(0)$ | $A_1(0)$ | $A_2(0)$ |
|---|---|---|---|---|---|
| $\Delta b = 0, \Delta C = -1, \Delta S = -1$ | $B_c \to B_s^*$ | $5.19^{+0.08}_{-0.11}$ | $0.57^{+0.02}_{-0.03}$ | $0.79^{+0.01}_{-0.02}$ | $3.24^{+0.04}_{-0.09}$ |
| | $B_c \to B^*$ | $4.77^{+0.07}_{-0.10}$ | $0.42^{+0.02}_{-0.02}$ | $0.63^{+0.01}_{-0.01}$ | $2.74^{+0.04}_{-0.07}$ |
| $\Delta b = 1, \Delta C = 0, \Delta S = -1$ | $B_c \to D^*$ | $0.16^{+0.02}_{-0.02}$ | $0.081^{+0.07}_{-0.08}$ | $0.095^{+0.013}_{-0.015}$ | $0.11^{+0.01}_{-0.02}$ |
| | $B_c \to D_s^*$ | $0.29^{+0.02}_{-0.03}$ | $0.16^{+0.01}_{-0.01}$ | $0.18^{+0.01}_{-0.02}$ | $0.20^{+0.02}_{-0.03}$ |
| $\Delta b = 1, \Delta C = 1, \Delta S = 0$ | $B_c \to J/\psi(c\bar{c})$ | $0.91^{+0.04}_{-0.05}$ | $0.58^{+0.01}_{-0.03}$ | $0.63^{+0.03}_{-0.03}$ | $0.74^{+0.05}_{-0.06}$ |

**Tables IV. Branching ratios (in $\frac{\tau_{B_c}(\text{ps})}{0.46}$ %) of $B_c \to PV$ decays (Errors shown here are due to uncertainty in $m_{B_c^*} - m_{B_c}$).**

| Decays | This Work | | [4] | [5] | [6] | [7] | [8] | [9] |
|---|---|---|---|---|---|---|---|---|
| | Br. (%) ($\omega$ = 0.40 GeV) | Br. (%) (using flavor dependent $\omega$) | | | | | | |
| $\Delta b = 0, \Delta C = -1, \Delta S = -1$ | | | | | | | | |
| $B_c^+ \to \pi^+ B_s^{*0}$ | 1.91 | $4.37^{+0.37}_{-0.33}$ | 7.33 | 1.78 | 3.95 | 5.73 | 2.37 | 1.39 |
| $B_c^+ \to B_s^0 \rho^+$ | 2.75 | $7.00^{+0.60}_{-0.50}$ | 8.10 | 1.55 | 12.22 | 4.97 | 2.60 | 4.35 |
| $B_c^+ \to B^+ \overline{K}^{*0}$ | 0.38 | $0.72^{+0.15}_{-0.00}$ | 1.14 | 0.23 | --- | 1.24 | 0.29 | 0.85 |
| $B_c^+ \to \overline{K}^0 B^{*+}$ | 0.38 | $0.80^{+0.00}_{-0.14}$ | 4.23 | 0.28 | --- | 1.33 | 0.23 | 0.44 |
| $\Delta b = 1, \Delta C = 1, \Delta S = 0$ | | | | | | | | |
| $B_c^- \to D^0 D^{*-}$ | $2.69 \times 10^{-5}$ | $6.57^{+1.21}_{-1.26} \times 10^{-4}$ | $8.50 \times 10^{-3}$ | --- | $6.89 \times 10^{-3}$ | $3.1 \times 10^{-3}$ | $1.50 \times 10^{-3}$ | $1.36 \times 10^{-3}$ |
| $B_c^- \to D^- D^{*0}$ | $3.53 \times 10^{-5}$ | $8.34^{+1.52}_{-1.56} \times 10^{-4}$ | 0.01 | --- | $6.14 \times 10^{-4}$ | $3.3 \times 10^{-3}$ | $6.60 \times 10^{-3}$ | $3.60 \times 10^{-3}$ |
| $B_c^- \to \eta_c \rho^-$ | 0.04 | $0.39^{+0.01}_{-0.03}$ | 0.41 | 0.20 | 0.07 | 0.48 | 0.43 | 0.33 |
| $B_c^- \to \pi^- J/\psi$ | 0.01 | $0.13^{+0.01}_{-0.01}$ | 0.13 | 0.06 | 0.13 | 0.17 | 0.17 | 0.11 |
| $\Delta b = 1, \Delta C = 0, \Delta S = -1$ | | | | | | | | |
| $B_c^- \to \eta_c D_s^{*-}$ | 0.04 | $0.31^{+0.01}_{-0.01}$ | 0.22 | --- | 0.48 | 0.03 | 0.33 | 0.20 |
| $B_c^- \to D_s^- J/\psi$ | 0.03 | $0.28^{+0.01}_{-0.01}$ | 0.13 | --- | 0.33 | 0.03 | 0.31 | 0.13 |

**Table V.** $|\psi(0)|^2$ and $\omega$ for vector and pseudoscalar mesons (Errors shown here are due to uncertainty in $m_{B_c^*} - m_{B_c}$).

| Meson | $|\psi(0)|^2$ (in GeV$^3$) | Parameter $\omega$ (in GeV) |
|---|---|---|
| $\rho(\pi)$ | 0.011 | 0.33 |
| $K^*(K)$ | 0.011 | 0.33 |
| $D^*(D)$ | 0.026 | 0.43 |
| $D_s^*(D_s)$ | 0.041 | 0.51 |
| $J/\psi(\eta_c)$ | 0.115 | 0.71 |
| $B^*(B)$ | 0.033 | 0.47 |
| $B_s^*(B_s)$ | 0.053 | 0.55 |
| $B_c$ | $0.281^{+0.077}_{-0.060}$ | $0.96^{+0.08}_{-0.07}$ |